\documentclass[prd,amsmath,amssymb,showpacs]{revtex4}
\usepackage{graphicx}
\usepackage{color}
\usepackage[colorlinks=true,linkcolor=red]{hyperref}
\usepackage{hyperref}

\begin{document}
\hfill{ITEP-TH-38/06}
\vspace{5mm}
\title{Hawking temperature in the tunneling picture}
\author{Emil T.Akhmedov}
\email{akhmedov@itep.ru} 
\affiliation{Moscow, B.Cheremushkinskaya, 25, ITEP, Russia 117218}
\author{Valeria Akhmedova}
\email{lera@itep.ru} \affiliation{Physics Department, CSU Fresno, Fresno, CA 93740-8031}
\author{Douglas Singleton}
\email{dougs@csufresno.edu} \affiliation{Physics Department, CSU Fresno,
Fresno, CA 93740-8031}

\date{\today}

\begin{abstract}
We examine Hawking radiation from a Schwarzschild black hole in
several reference frames using the quasi-classical tunneling
picture. It is shown that when one uses, $\Gamma \propto \exp(Im [\oint ~ p ~ dr])$,
rather than, $\Gamma \propto \exp(2 Im [\int ~ p ~ dr])$, 
for the tunneling probability/decay rate one obtains twice the
original Hawking temperature. The former expression for $\Gamma$ is argued
to be correct since $\oint ~ p ~ dr$ is invariant under canonical 
transformations, while $\int ~ p ~ dr$ is not. Thus, either the tunneling
methods of calculating Hawking radiation are suspect or the Hawking 
temperature is twice that originally calculated.
\end{abstract}

\pacs{04.62.+v, 04.70.Dy, 03.65.Xp}

\maketitle


\section{Introduction}

The original derivation of Hawking radiation from a Schwarzschild black hole
\cite{hawking} was done with the methods of quantum field theory 
in a curved background. The physical picture given for this
process was in terms of particles tunneling through the horizon which was not
directly connectable with the field theory derivation. Relatively
recently this physical picture has been given support by quasi-classical
tunneling calculations \cite{ volovik} \cite{parikh} \cite{padmanabhan} \cite{vagenas} 
\cite{zerbini} \cite{mann}. The present approach, where we use the Hamilton-Jacobi
equations to study black hole radiation, is that used in \cite{padmanabhan}
\cite{zerbini}. All the above works are more directly connected to the physical picture
of Hawking radiation as tunneling, and are more transparent and less
complex than the field theory derivation. In a recent work \cite{akhmedov}
we studied Hawking like radiation in various gravitation backgrounds using
the quasi-classical tunneling approach and the Hamilton-Jacobi equations. 
In the case of Schwarzschild black holes we found the physically questionable
result that the Hawking temperature apparently depended on the type of
coordinates used ({\it e.g.} Schwarzschild, Painlev{\'e}, isotropic). Further in
\cite{chowdhury} it was argued that since the quantity
$\int  p  dr$ is not invariant under canonical transformations, that the tunneling
probability or decay rate, $\Gamma \propto \exp(2 Im [\int ~ p ~ dr])$, is not a proper
observable in this particular case of tunneling through the black hole horizon. 
In this paper we will discuss these issues
and suggest resolutions to these problems with the tunneling calculations
of Hawking radiation. We give arguments that if the tunneling
calculations are correct then the temperature of the Hawking radiation
should be twice as large as originally calculated. 

\section{Hamilton-Jacobi equation approach to black hole radiation via tunneling}
 
To study the thermal radiation given off in some gravitational background
we consider the Klein--Gordon equation in a curved background:
\begin{equation} 
\left[- \frac{\hbar^2}{\sqrt{-g}} \partial_\mu g^{\mu\nu} \sqrt{-g} \partial_\nu\, 
+ m^2\right] \phi = 0. \label{KG}
\end{equation} 
The signature of the metric is (-1,1,1,1)
and $ds^2 = g_{\mu\nu}(x) \, dx^\mu \, dx^{\nu}$, $g_{\mu\nu} \,
g^{\nu\alpha} = {\delta_{\mu}}^\alpha$. In this paper we do not take into 
account the back--reaction of gravity on the quantum fluctuations of the scalar field.

We are looking for the solutions of \eqref{KG} having the form: 
$\phi(x) \propto \exp\left\{-\frac{\rm i}{\hbar} \, S(x) +
\dots \right\}$. 
Inserting this into \eqref{KG} and taking the limit $\hbar \to 0$ we find to
order $\hbar^0$ the following equation
\begin{equation} 
g^{\mu\nu} \, \partial_\mu S \, \partial_\nu S + m^2 = 0,\label{HJ}
\end{equation} 
which is the relativistic Hamiltonian--Jacobi equation
for the classical action of a relativistic particle in a curved
background. The condition under which our approximation is
valid is worked out in \cite{landau}.

Since the Schwarzschild metric is stationary it has
a time--like Killing vectors. Thus, we will look for
particle--like solutions of \eqref{HJ} which behave as
$S = E\, t + S_0(\vec{x})$ , 
where $x_\mu = (t,\vec{x})$ and $E$ is the energy of the
particle. The wave function for such solutions behaves 
as $\phi \propto e^{-\frac{\rm i}{\hbar} \, E\, t}$ 
and corresponds to a state with definite energy. It is these states 
which are observed by detectors. 

If the solution $S_0(\vec{x})$ of \eqref{HJ} has a
non--zero imaginary part for some particle trajectory this implies
that the gravitational background in question is unstable with respect 
to radiation of such definite energy states. In this
case the wave function behaves as:
$\phi \propto e^{- \frac{1}{\hbar} \, {\rm Im} S_0}$,
which describes tunneling of the particle through the gravitational
barrier. This leads to the decay of the background with the rate
given by $\Gamma \propto |\phi|^2 \propto e^{-\frac{2}{\hbar} \, {\rm
Im} S_0 }$.

In Schwarzschild coordinates an uncharged, non-rotating black hole with mass $M$
has a metric of the form 
\begin{equation} 
ds^2 = - \left(1 - \frac{2M}{r}\right) \, dt^2 +
\frac{dr^2}{\left(1 - \frac{2M}{r} \right)} + r^2 d\Omega^2 . \label{schwarz}
\end{equation}
For this metric and for radial trajectories which are independent of
$\theta , \varphi$ the Hamilton--Jacobi equation becomes
\begin{equation} 
- \frac{1}{\left(1 - \frac{2M}{r}\right)}\, 
\left(\frac{\partial S}{\partial t}\right)^2 + \left(1 - \frac{2M}{r}\right)\,
\left(\frac{\partial S}{\partial r}\right)^2 + m^2 = 0.
\end{equation} 
For the definite energy state we obtain
\begin{equation} 
- \frac{E^2}{\left(1 - \frac{2M}{r}\right)} + \left(1 -
\frac{2M}{r}\right)\, \left(\frac{dS_0}{dr}\right)^2 + m^2 =
0.\label{diff}
\end{equation}
Despite the fact that the Schwarzschild metric has two disjoint parts separated
by $r=2M$, we can nevertheless consider solutions of \eqref{diff} in these two regions and 
glue them by going around the pole in the complex $r$-plane.  
The solution is
\begin{equation} S_0 = \pm \int_0^{+\infty} \frac{dr}{\left(1 -
\frac{2M}{r}\right)}\, \sqrt{E^2 - m^2 \left(1 -
\frac{2M}{r}\right)},\label{intsch}
\end{equation} 
where the limits of integration are chosen such that the particle goes through
the horizon $r=2M$. We focus on the integration through $r=2M$ since this is exactly 
where the complex part of $S_0$ comes from. The $+ (-)$ sign in front of this integral indicates
that the particle is ingoing (outgoing). Thus both incoming and outgoing particles 
face barriers as should be expected for a barrier penetration/tunneling problem. However
classically this is odd since one expects a classical particle to face a barrier
only when it is outgoing not incoming. This point will be discussed in detail later, but
note that for virtual particle pairs the positive energy component must tunnel
out from the horizon, while the negative energy component must tunnel in through the
horizon. 

Because there is a pole at $r=2M$ along the path of 
integration the integral will just be the Cauchy principle 
value. The imaginary part of the principle value of \eqref{intsch}
is given by the contour integral over a
small half--loop going around the pole from below from
left to right. This choice of the contour seem to correspond to
the small ``trajectory'' of the particle just crossing the horizon.
It is worth mentioning at this point that there is no
real trajectory of a classical particle corresponding to this
contour even in Euclidian time. This observation makes the whole quasi-classical 
approach questionable. However, in a moment we will give independent arguments
why this is nevertheless correct.

First, let us explicitly take the imaginary part of the principle value. We 
make the change of variables $r-2M = \epsilon\, e^{{\rm i}\, \theta}$. Then

\begin{equation} 
{\rm Im} S_0 = \pm \lim_{\epsilon \to 0} \int_{\pi}^{2\, \pi}
\frac{\left(2M + \epsilon\, e^{{\rm i}\, \theta}\right) \,
\epsilon\, e^{{\rm i}\, \theta} \, {\rm i}\, d\theta}{\epsilon \,
e^{{\rm i}\, \theta}}\, \sqrt{E^2 - m^2 \left(1 - \frac{2M}{2M +
\epsilon\, e^{{\rm i}\, \theta}}\right)} = \pm 2\,\pi\, M\,
E.\label{Hawef}
\end{equation} 
Using this result for ${\rm Im} S_0$ for the outgoing particle 
the decay rate of the black hole is
$\Gamma \propto e^{\frac{-4\,\pi\,M\,E}{\hbar}}$. 
From the above expression for $\Gamma$ one sees that it is
just the Boltzman weight with the temperature $T= \hbar /4\,\pi\,M$.

Before continuing our quasi-classical analysis in other frames a comment
on the form of the decay rate is in order. The behavior of the decay rate -- 
$\Gamma\propto e^{-\frac{4\,\pi\,M\,E}{\hbar}}$ -- seems strange from the
point of view of quasi-classics. The tunneling rate decreases with 
increasing energy, which is contrary to the ``quasi-classical'' intuition.
However, there is an obvious explanation for this behavior: the greater
the energy of a particle the stronger it is attracted to the gravitating body (black hole),
hence, the harder it is for it to escape.

Now let us come back to the imaginary part of the action. We have obtained the 
temperature $T= \hbar /4\,\pi\,M$ which is twice the temperature originally calculated by 
Hawking. We now look further into this discrepancy. In \cite{parikh} the tunneling calculation 
was performed not in the Schwarzschild frame of \eqref{schwarz} but in the Painlev{\'e} frame
\begin{equation} 
ds^2 = - \left(1-\frac{2M}{r}\right)\, dt^2 +
2\sqrt{\frac{2M}{r}}\, dr\,dt + dr^2 + r^2 \, d\Omega^2  , \label{pmetric}
\end{equation} 
(The use of the Painlev{\'e} frame to study tunneling in thin film $^3 He$ 
black hole analog systems was first employed in \cite{volovik})
The Painlev{\'e} frame is obtained from the Schwarzschild frame by making the following 
transformation of the {\it time} coordinate
\begin{equation} 
dt' = dt + \frac{\sqrt{\frac{2M}{r}} \, dr}{1-\frac{2\,
M}{r}},\quad r' = r, \quad \Omega' = \Omega . \label{chan}
\end{equation} 
This metric is regular (i.e. does not have the horizon for the {\it incoming}
particles) at $r=2M$. However the notion of time is changed with respect to the
Schwarzschild coordinates, so that the
Hamiltonian--Jacobi equation for the definite energy state becomes
\begin{equation}
- E^2 + \left(1- \frac{2\, M}{r}\right)\,
\left(\frac{dS_0}{dr}\right)^2 + 2\,\sqrt{\frac{2\, M}{r}}\,
E\,\frac{dS_0}{dr} + m^2 = 0.
\end{equation} 
The solution of this equation is
\begin{equation} S_0 = - \int_{C}
\frac{dr}{1-\frac{2M}{r}}\,\sqrt{\frac{2M}{r}}\, E \pm \int_{C}
\frac{dr}{1-\frac{2M}{r}}\,\sqrt{E^2 - m^2 \left(1 -
\frac{2M}{r}\right)}.\label{ft}
\end{equation} 
This result can not be obtained from \eqref{intsch} via a change of 
integration variables because the transformation \eqref{chan} does affect 
the time--like Killing vector. One can see that \eqref{ft} differs from \eqref{intsch}
by the first term. The first term in \eqref{ft}
arises from the coordinate change in \eqref{chan}, since
\begin{equation} 
\int E dt + S_0 = \int E dt' - \int
\frac{dr}{1-\frac{2M}{r}}\,\sqrt{\frac{2M}{r}}\, E + S_0, 
\end{equation}
where $S_0$ is given by \eqref{intsch}. Physically this 
new coordinate system corresponds to $r$-dependent, singular shift of
the initial time. 

If we choose the plus sign in \eqref{ft} (which corresponds 
to an incoming particle) and the contour $C$ as in
\eqref{Hawef}, we find ${\rm Im} S_0 =0$ since the first and
second terms in \eqref{ft} have the same magnitude. Thus incoming
particles do not see a barrier or horizon. If on the other hand 
we choose the minus sign in \eqref{ft} and the contour $C$ as
before the result is
\begin{equation} 
{\rm Im} S_0 = - 4\,\pi\, M\, E.\label{Haw1}
\end{equation} 
This is twice the result of \eqref{Hawef} because the first integral in
\eqref{ft} gives the same contribution to the complex part of $S_0$
as the second one. Thus in this frame, if one uses $\Gamma \propto |\phi|^2 \propto e^{-\frac{2}{\hbar} \, {\rm
Im} S_0 }$ for the decay rate, one apparently recovers the original
temperature of the black hole as calculated in \cite{hawking} namely
$T=\hbar / 8 \pi M$. Note that in Hawking's original derivation he also did
not use the Schwarzschild frame but rather used a 
frame where the time $t'$ was related to the Schwarzschild time via
$dt' = dt + dr/(1 - \frac{2M}{r})$. As in the transformation from 
Schwarzschild frame to Painlev{\'e} frame, this involves a shifting of
the time coordinate. 

In \cite{zerbini} it was argued that this apparent disagreement between the
results in the Schwarzschild frame and in the Panilev{\'e}
frame (or in the frame used by Hawking in \cite{hawking} mentioned in
the previous sentence) 
was a result of the bad behavior of the Schwarzschild coordinates
at $r=2M$. By making a change of {\it spatial} variables to a frame 
where the coordinates were better behaved at the horizon it was claimed 
one would recover the original result of Hawking for the temperature.
As a particular example one could consider isotropic coordinates which 
are obtained from the Schwarzschild coordinates via the change of variables
\cite{zerbini}
\begin{equation}
\label{iso-coor}
r=\rho \left( 1 + \frac{M}{2 \rho} \right) ^2 .
\end{equation}
With this the Schwarzschild metric \eqref{schwarz} becomes
\begin{equation}
\label{iso-metric}
ds^2 = - \left( \frac{\rho - \frac{M}{2}}{\rho + \frac{M}{2}} \right)^2 dt^2
+\left( \frac{\rho + \frac{M}{2}}{\rho} \right)^4 \left( d \rho ^2 + \rho ^2 d \Omega ^2 \right).
\end{equation}
Now instead of \eqref{intsch} we find
\begin{equation} S_0 = \pm \int \frac{(\rho + \frac{M}{2})^3}{(\rho -\frac{M}{2}) \rho^2} 
\sqrt{E^2 - m^2 \left( \frac{\rho -\frac{M}{2}}{\rho + \frac{M}{2}}\right)^2} ~ d \rho,\label{intiso}
\end{equation} 
If one does the contour integration in the same manner as in \eqref{Hawef} by making a
semi-circular contour one apparently finds that Im $S_0 = \pm 4 \pi M E$ \cite{zerbini}. 
However there is a subtle point: one must also
deform the contour from \eqref{Hawef} using \eqref{iso-coor} and when this is done the semi-circular 
contour of \eqref{Hawef} gets transformed into a quarter circle so that one gets $i \frac{\pi}{2} Residue$
\footnote{It is this choice of the contour, rather than any 
other one, which corresponds to going from
inside to outside the horizon.}  rather than $i \pi Residue$.
One could already guess this because from \eqref{iso-coor} $\rho \simeq
\sqrt{r}$ which for the contour in \eqref{Hawef} means the semi-circle contour becomes a 
quarter circle. In detail
\begin{equation}
r=2M-\epsilon e ^{i \theta} = \rho + M +\frac{M^2}{4 \rho} \rightarrow
\rho = \frac{1}{2} \left( M + \epsilon e^{i \theta}
\pm  (2 M + \epsilon e^{i \theta} ) \sqrt{\epsilon} e^{i \theta /2} \right)
\end{equation}
The leading order in epsilon is now $\sqrt{\epsilon}$ so in the limit $\epsilon
\rightarrow 0$ we find from the above equation $\rho -\frac{M}{2} = M \sqrt{\epsilon}
e^{i \theta /2}$ instead of $r- 2M = \epsilon e^{i \theta}$. Since $e^{i \theta}$
becomes $e^{i \theta /2}$, one sees that the semi-circle
contour of the Schwarzschild frame gets transformed into a quarter circle in the isotropic 
coordinate frame so that the result of integrating \eqref{intiso} is $i \frac{\pi}{2} Residue$ and
we find again ${\rm Im} S_0 = \pm 2 \pi M E$.

In \cite{zerbini} general arguments are given that one should work with the proper spatial 
distance as defined by
\begin{equation}
\label{proper}
d\sigma ^2 = \frac{dr^2}{B(r)} + r^2 d\Omega^2 \rightarrow
\sigma = \int \frac{dr}{\sqrt{B(r)}}
\end{equation}
where $B(r) = 1- \frac{2 M}{r}$. In the last step we are considering only the radial part 
or the s-wave contribution to the tunneling. In \cite{zerbini} by considering the
near horizon approximation (i.e. $B(r) = B'(r=2M) (r- 2M) + ...$) one finds
\begin{equation}
\label{near-h}
\sigma = 2 \sqrt{2 M} \sqrt{r- 2M} + ...
\end{equation}
From this one sees that in general the contour from \eqref{Hawef} defined
via $r-2M = \epsilon\, e^{{\rm i}\, \theta}$ will always be transformed from
a semi-circle to a quarter circle because of the square root in 
\eqref{near-h}. Thus any coordinate transformation of \eqref{Hawef} involving
only {\it spatial} coordinates will yield the same result for the 
temperature as \eqref{Hawef}. This can also be seen from the point that
making a coordinate transformation involving the spatial coordinates is
just a change of integration variables and should not change the result. 

Thus, if one uses Schwarzschild coordinates or any coordinates
related to them via a transformation of {\it spatial}
coordinates one gets twice the original Hawking temperature, while transformations
involving the {\it time} coordinate {\it appear} to give the original Hawking result.
How can one reconcile these various results? In fact, a detector will only measure
one or the other of these temperatures via say the rate of flipped and un--flipped
spins. 

We now show that even in the Painlev{\'e} frame one obtains 
twice the Hawking temperature if one
takes the proper exponent in the expression for $\Gamma$.
In \cite{chowdhury} the tunneling approach to Hawking radiation was
criticized based in the fact that $2{\rm Im} S_0 = 2{\rm Im}\int  p  dr$ is not
invariant under canonical transformations and thus 
$\Gamma \propto \exp(2 {\rm Im} S_0) = \exp(2 {\rm Im} [\int  p dr])$ is not a proper observable;
one could change $\Gamma$ by making a canonical transformation. In \cite{chowdhury} it is argued that since
the closed contour integral, $\oint  p  dr$, is invariant under
canonical transformations one should take for the decay rate
$\Gamma \propto \exp({\rm Im} [\oint  p  dr])$. The relationship between
the two expressions can be seen by considering a closed path that
goes from $r=r_i$, which in just inside the horizon, to $r=r_o$ just
outside the horizon
\begin{equation}
\label{path}
\oint p dr = \int _{r_i} ^{r_o} p_{out} dr + \int _{r_o} ^{r_i} p_{in} dr
\end{equation}      
The points, $r_i , r_o$ are chosen to straddle the horizon since this is where
the imaginary part of $S_0$ comes from. If, as suggested in \cite{chowdhury},
one takes \eqref{path} to define the exponent in $\Gamma$ then one finds that
all three frames considered -- Schwarzschild, Painlev{\'e} and isotropic --
yield the same result namely Im$\oint p dr = 4 \pi M E$ which then in all cases 
gives $\Gamma \propto e^{\frac{-4\,\pi\,M\,E}{\hbar}}$ and a temperature of,
$T= \hbar /4\,\pi\,M$, twice the Hawking temperature. In the Schwarzschild and
isotropic coordinates the $p_{out}$ and $p_{in}$ have equal magnitude, but 
opposite signs -- the $+(-)$ signs correspond to $p_{out} (p_{in})$ in 
\eqref{intsch} \eqref{intiso}. For these two coordinate frames one can see
why $\oint p dr = 2 \int _{r_i} ^{r_o} p_{out} dr = 2 \int _{r_o} ^{r_i} p_{in} dr$. 
In the Painlev{\'e} coordinates the entire
contribution comes from $p_{out}$, since for $p_{in}$ the two contributions 
in \eqref{ft} cancel. For the Painlev{\'e} coordinates
$\oint p dr \ne 2 \int _{r_i} ^{r_o} p_{out} dr \ne 2 \int _{r_o} ^{r_i} p_{in} dr$.
The fact that one gets a contribution from both incoming
and outgoing particles when one uses Schwarzschild or isotropic coordinates
is in accordance with the idea that in a proper tunneling problem one should face a barrier regardless
of whether one moves from left (inside) to right (outside) or from right (outside)
to left (inside) across the barrier. However from the classical point of view this
appears odd since a classical particle can easily cross the horizon going inward. It is
only crossing the horizon in the outward direction that is forbidden for a classical
particle. However in the tunneling picture one
is looking at virtual pairs of negative-positive energy particles which are fluctuating out
of the vacuum near the horizon. For such a virtual pair just inside the horizon, the positive 
energy part must tunnel out of the horizon with the negative energy component going
inward. Just outside the horizon it is the negative energy component 
which must tunnel through the horizon with the positive energy component going outward. 
Thus contrary to classical intuition the horizon represents a two way
barrier when one considers virtual particle pairs.

We give two further arguments supporting our conclusions.
The calculation of the Unruh temperature in our previous paper \cite{akhmedov}
is in accordance with the considerations of the previous paragraph.
Similar to the Schwarzschild metric,  the Rindler 
metric has a horizon for the particles going both ways
through its singular point. Also, the corresponding decay rate is given by $\log \Gamma \propto
{\rm Im} \, \oint p \, dr$, which gives the correct Unruh temperature $T = a/2\pi$,
where $a$ is the acceleration. The second argument follows by considering the 
scattering problem in the black hole background (see e.g. \cite{PhysRept}). 
The Klein--Gordon equation in the Schwarzschild background:

\begin{equation}
\left[-\frac{1}{\left(1 - \frac{2M}{r}\right)}\, \frac{\partial^2}{\partial t^2} + \frac{1}{r^2}\,
\frac{\partial}{\partial r}\, r^2 \left(1 - \frac{2M}{r}\right)\, \frac{\partial}{\partial r} + \frac{1}{r^2}\,
\Delta(\theta,\varphi) - m^2\right]\,\phi = 0, 
\end{equation} 
where $\Delta(\theta,\varphi)$ is the angular part of the Laplacian.
We would like to find the behavior of the constant energy solution ($\phi = e^{i\, E\, t}\, \phi_E$) 
in the vicinity of the horizon $r=2M$. Changing the variables to $z = r - 2M$, 
the Klein--Gordon equation approximately reduces to:

\begin{equation}
\left[\left(2ME\right)^2 + z\,\frac{d}{dz} \,z\,\frac{d}{dz}\right] \phi_E = 0.
\end{equation} 
Thus, near the horizon the wave function behaves as follows:

\begin{equation}
\phi_E \propto z^{{\rm i} \, 2\, M\, E} \propto \left(r-2M\right)^{{\rm i} \, 2\, M\, E}. 
\end{equation} 
The power in this formula is directly related to the imaginary part
of the action which we considered above.
To solve the scattering problem in question we have to connect solutions inside
and outside the horizon \cite{PhysRept}. This is achieved by going to the complex $r$ plane
and taking the above mentioned contour around $r=2M$. In this way 
we obtain a damping factor of $e^{-2\pi\,M\,E}$ \cite{PhysRept} which 
corresponds to a temperature of $T= \hbar /4\,\pi\,M$.

As a final comment tunneling calculations in general (if one has canonical invariance)
seem to give a factor of two greater temperature as compared quantum field theory 
calculations. In \cite{brout} the tunneling calculation of the Gibbons-Hawking
temperature of a de Sitter spacetime gave a temperature twice that of the
quantum field theory calculation, but their explanation of the discrepancy in question 
is not applicable to our case. The Unruh effect is only case we have examined thus far 
where the tunneling and other calculations agree. 

\section{Conclusions}

We have shown that by requiring canonical invariance (i.e. 
using quantity $\oint p dr$ to define the exponent
of the decay rate, $\Gamma$) the tunneling calculations of black hole radiation
all give a temperature that is twice that was originally calculated by Hawking.
There are two possible conclusions: 
(i) the tunneling calculations are not correct in detail or (ii)
the Hawking temperature is twice as large as the original 
calculations suggest. (A third possibility has been suggested to us 
\cite{parikh2}: that canonical invariance is not
a requirement in the tunneling calculations as applied to black holes. We can find
no argument why this should be and thus do not take this option into account).
We now discuss the pros and cons of these two possibilities. 
If one takes the second possibility as correct 
then the tunneling calculations, although corresponding to the
physical picture of Hawking radiation as tunneling through the horizon, are
only a heuristic guide which in detailed calculations do not give the correct numerical 
factor for the temperature. However, the standard calculation of the
Hawking temperature is not airtight; there are open questions: the appearance of trans-Planckian
energies at intermediate stages of the calculation and the unknown effect of
quantum gravitational corrections \cite{gibbons} \cite{jacob}. Reference \cite{helfer} gives an
excellent discussion of these issues and other open issues in the standard derivation of
Hawking radiation. Finally we note that in the original calculation of Hawking radiation,
and in almost all subsequent calculations, the Schwarzschild frame is not used, but rather 
the frame is used where the time $t'$ is related to the Schwarzschild time via
$dt' = dt + dr/(1 - \frac{2M}{r})$. This is similar to the transformation
\eqref{chan} that takes one from Schwarzschild to Painlev{\'e} coordinates. Changing the
time coordinate in this way changes the quantization of the fields, and could lead to
subtleties such as those that arose when one did the tunneling calculation in Painlev{\'e} coordinates. 
In fact, in different frames (Schwarzschild or isotropic)
even using Hawking's approach one should use a different basis of Fourier harmonics,
which could then lead to different results for the black hole temperature.
There was a previous suggestion that the temperature of Hawking radiation should be 
twice the originally calculated value \cite{thooft}. In \cite{thooft} it was argued
that any horizon splits the Universe into two parts, both of which give a contribution 
to the thermal radiation. This is similar to the tunneling picture we sketched at
the end of the last section: virtual particles just inside the horizon {\it and} just outside the horzion, 
both contribute to the tunneling and thus to the thermal radiation. 

We close by making some observations on the effects of having a Hawking 
temperature twice as large as originally calculated. Astrophysically 
doubling the Hawking temperature would have no 
experimental significance, since for astronomical black holes Hawking
radiation is too weak to detect. However, in some extra dimensional models, 
gravity becomes strongly coupled at the TeV scale with the possibility of creating micro
black holes at the LHC \cite{giddings}. These micro black holes would rapidly decay via Hawking 
radiation and in this case the factor of two difference would be noticeable. Another 
area where having the Hawking temperature twice as large would make a difference is
in the entropy of a black holes. For a non-rotating, chargeless, black hole the $1^{st}$
law of black hole thermodynamics reads (taking $G=1, c=1, k_B=1$)
\begin{equation}
\label{entropy}
dM = \frac{\kappa}{8 \pi} d A \qquad {\rm with} \qquad \kappa =\frac{1}{4 M}.
\end{equation} 
Comparing with the ordinary $1^{st}$ law ($dU = T ~ dS$) and assuming that the Hawking
temperature is $T=\hbar / 8 \pi M$ then leads to the Hawking-Bekenstein \cite{bekenstein} 
expression for black hole entropy: $dS= dA / 4 \hbar \rightarrow S=A/ 4 \hbar$ where $A$ is the area
of the horizon. One of the theoretical tests of quantum gravity theories such
as string theory or loop quantum gravity is to calculate the entropy of a black hole 
by counting the microscopic states. In certain special situations (i.e. extremal, supersymmetric
black holes) there have been string theory calculations \cite{string} of black hole entropy. 
A review of this topic can be found in \cite{Akhmedov1}.
In loop quantum gravity there have also been calculations of black hole entropy \cite{loop}.
The goal is to reproduce the result $S= A / 4 \hbar$. However if as indicated
by the tunneling calculations the temperature is really $T=\hbar / 4 \pi M$ this would
imply $S=A/ 8 \hbar$ which would be important for quantum gravity calculations which
try to reproduce the black hole entropy by counting microscopic states.  

\section{Acknowledgments}

AET would like to thank S.Dubovski, S.Sibiryakov, V.Rubakov and especially
A.Morozov and J.Bjorken for illuminating discussions. This work
supported by the following grants: RFBR 04-02-16880 and the Grant from the President 
of Russian Federation for support of scientific schools, and 
a CSU Fresno International Activities Grant.


\end{document}